\documentclass[11pt]{article}

\usepackage{endnotes}

\title{Exploring Connections Between Cosmos \& Mind\\ Through Six Interactive Art Installations in\\ ``As Above As Below''}
\author{Mark Neyrinck\endnotemark[1], Tamira Elul\endnotemark[2], Michael Silver\endnotemark[3], Esther Mallouh\endnotemark[4], \and
Miguel Arag\'{o}n-Calvo\endnotemark[5], Sarah Banducci, Cory Bloyd,
Thea Boodhoo, Benedikt Diemer\endnotemark[6], \and
Bridget Falck\endnotemark[7], Dan Feldman\endnotemark[8], Yoon Chung Han\endnotemark[9], Jeffrey Kruk\endnotemark[10], Soo Jung Kwak, \and
Yagiz Mungan, Miguel Novelo\endnotemark[11], Rushi Patel, Purin Phanichphant\endnotemark[12], Joel Primack\endnotemark[13], \and
Olaf Sporns\endnotemark[14], Forest Stearns\endnotemark[15], Anastasia Victor\endnotemark[16],
David Weinberg\endnotemark[17], Natalie M. Zahr\endnotemark[18]}
\date{SciArt Magazine, February 2020 \url{https://www.sciartmagazine.com/collaboration-as-above-as-below.html}}

\usepackage{natbib}
\usepackage{graphicx}
\usepackage{patchcmd}

\usepackage{natbib}
\usepackage{float}
\usepackage{hyperref}
\usepackage[symbol]{footmisc}
\usepackage[T1]{fontenc}

\usepackage{caption}
\usepackage[margin=0.85in,bottom=1in,top=1in]{geometry}

\usepackage{titlesec}
\titleformat{\section}
  {\normalfont\sffamily\Large\bfseries}
  {\thesection}{1em}{}
\titleformat{\subsection}
  {\normalfont\sffamily\large\bfseries\itshape}
  {\thesection}{1em}{}
  
\usepackage{fancyhdr}
\newcommand{\longlinetwo}{\underline{\hspace{2cm}}}
\newcommand{\longlineone}{\underline{\hspace{1cm}}}

\begin{document}

\sffamily
\setlength{\parskip}{8pt}

\maketitle

\endnotetext[1]{Ikerbasque Fellow; University of the Basque Country, Bilbao; Donostia International Physics Center, San Sebasti\'{a}n, Spain}
\endnotetext[2]{Touro University California}
\endnotetext[3]{Helen Wills Neuroscience Institute, School of Optometry, and Vision Science Graduate Group, University of California, Berkeley, CA 94720, USA}
\endnotetext[4]{Curator and organizer; Keen on Art}
\endnotetext[5]{UNAM, Ensenada, Mexico}
\endnotetext[6]{NHFP Einstein Fellow, Department of Astronomy, University of Maryland, College Park, MD 20742, USA}
\endnotetext[7]{Physics and Astronomy Dept, Johns Hopkins University}
\endnotetext[8]{Dept of Molecular \& Cell Biology, UC Berkeley}
\endnotetext[9]{San Jos\'{e} State University}
\endnotetext[10]{NASA Goddard}
\endnotetext[11]{Stanford University}
\endnotetext[12]{UC Berkeley}
\endnotetext[13]{Physics Department, University of California, Santa Cruz, Santa Cruz, CA 95064, USA}
\endnotetext[14]{Dept of psychological and Brain Sciences, Indiana University}
\endnotetext[15]{Google AI Quantum Artist in Residence}
\endnotetext[16]{Oculus}
\endnotetext[17]{Department of Astronomy, The Ohio State University, Columbus, OH 43210, USA}
\endnotetext[18]{SRI International}

\begin{abstract}
\normalsize Are there parallels between the furthest reaches of our universe, and the foundations of thought, awareness, perception, and emotion? What are the connections between the webs and structures that define both? What are the differences? ``As Above As Below'' was an exhibition that examined these questions. Conceptualized, curated, and produced by Esther Mallouh, it consisted of six artworks, each of them the product of a collaboration that included at least one artist, astrophysicist, and neuroscientist. The installations explored new parallels between intergalactic and neuronal networks through media such as digital projection, virtual reality, and interactive multimedia, and served to illustrate diverse collaboration practices and ways to communicate across very different fields.
\end{abstract}

A short video walkthrough of the exhibition is at \url{https://youtu.be/RM3gDjb3phU}, and a panel discussion at the opening, featuring many of the authors, is at \url{https://youtu.be/qNhtyvPz8yw}. See \url{https://asaboveasbelow.com/} for even more information about the project.

\section*{Introduction}
\begin{figure}
    \includegraphics[width=\columnwidth]{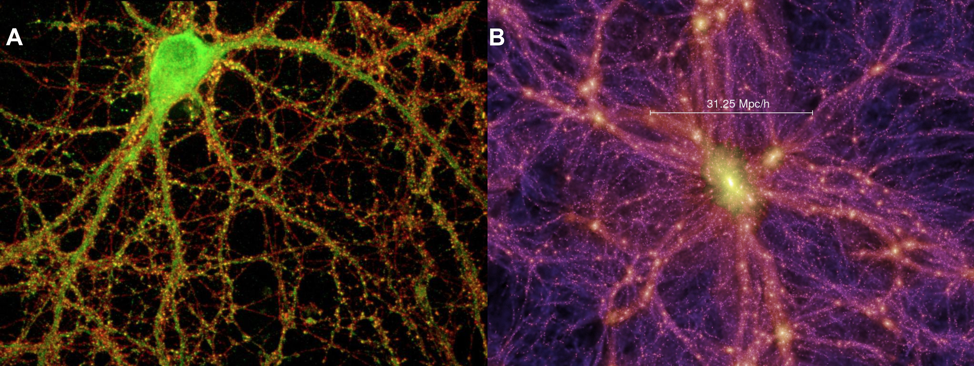}
  \caption{\small \textbf{A}. Hippocampal mouse neuron studded with synaptic connections (yellow), courtesy Lisa Boulanger, from \url{https://www.eurekalert.org/multimedia/pub/81261.php}. The green central cell body is $\sim10 \mu$m in diameter. \textbf{B}.\ Cosmic web \citep{mill}. Scale bar = 31.25 Mpc/$h$, or $1.4\times 10^{24}$ m. Juxtaposition inspired by \citet{Lima2009}.}
  \label{fig-1_2}
\end{figure}

The visual similarities between cosmic and neural webs are striking. Fig.\,\ref{fig-1_2}A shows a single mouse hippocampal neuron with an elaborate dendritic arbor stemming from a central cell body (large green blob); Fig.\,\ref{fig-1_2}B shows a cluster from a cosmological simulation, in which a whole galaxy like ours would be but a middling yellow dot. 

\vspace{-1mm}

However, many networks in the universe exhibit these kinds of visual similarities. For example, other biological systems, such as circulatory and respiratory networks in animals, and branching structures in plants and fungi, appear individually to have nearly maximal fractal dimension, and therefore look similar \citep{West2018}. These features are also evident in networks created by humans, such as traffic in road and air-travel systems, and non-spatial networks such as the Internet. See \citet{CoutinhoEtal2016} for more about the network-science properties of the cosmic web.
 
With so many types of similar spatial networks to pick from, why focus on cosmic and neural webs? One compelling reason is that they are each fascinating on their own: one is the network that is able to reflect on and study itself, and the scale of the other is nearly as big as the observable Universe. It is also curious that they are of similar complexity, with numbers of nodes that are within a couple of orders of magnitude of each other \citep{VazzaFeletti2017}.
 
The term ``cosmic web'' was first coined by \citet{BondEtal1996}. It refers to the network of filaments and walls of dark matter and gas that connect nearby galaxies, in the standard cosmological paradigm. The cosmic web exists on multiple spatial scales; there are also filaments of galaxies that connect clusters of galaxies. The cosmic web, and structural trussworks like spiderwebs and trees, also share a geometry that has been a source of artistic inspiration \citep{Ball2017}. To anthropomorphize: gravity acts like a haunted house explorer, clearing away flexible cobwebs from cosmic voids and gathering structural elements together into thicker filaments \citep{NeyrinckEtal2018}. The cosmic web moves at an aeonically slow pace in the expanding Universe, set in motion just after the Big Bang and nearly deterministically forming since then.

The cosmic web is not a pathway of communication -- the distances are far too great. But the whole point of the neural net is communication, on highways under constant construction and remodeling as the brain processes information and learns. In fact, interneuronal communication is likely to be the driving force in the evolution of neural webs. Like cosmic webs, neural webs exist on various scales: nervous systems are composed of neurons and clusters of neurons that exhibit network organization \citep{sporns2010networks}, dendrites and axons of individual neurons have complex branching structures (Fig.\,\ref{fig-1_2}A), and each neuron contains a molecular cytoskeleton that is made up of a network of microfilaments.
 
The principles underlying the development of specific neuron connectivity patterns in the neural web, hypothesized originally by the Spanish neuroanatomist Santiago Ramon y Cajal (and still relevant for theoretical and experimental neuroscientific studies today) are maximum efficiency of communication and meeting the metabolic challenge of making and supporting new connections \citep{cajal1999texture,wen2009maximization}. This leads to neurons that develop morphologies that approximate minimal spanning trees that contain a set of nodes that are branch points and/or synapse locations (Cuntz, 2012). Neurons have also inspired many artistic interpretations and visualizations \citep{Kim2015,PatelEtal2017}.
 
There are yet more differences between neural and cosmic webs. Neurons have objective physical membrane boundaries, while the boundaries of the filaments and walls of the cosmic web are subject to definition \citep[e.g.][]{LibeskindEtal2018}. Also, each neuron has a defined polarity (dendritic input and axonal output). In the cosmic web, galaxy clusters can locally look like neurons, but there is no analogous polarity, and no objective boundaries separating galaxy clusters. Another difference is that neurons have objective centers (somas, containing cell nuclei), but the Universe does not, nor do patches of the cosmic web. Any given region of the Universe on really large scales looks pretty much like any other large region of the Universe -- the large scale architecture of the observable Universe doesn't have any special regions. However, our brains are organized very differently -- vertebrate brains are networks with clearly identified specialized regions.

\section*{Art Installations}
\subsection*{\textit{Chamber of (In)finite Potential}}
An interactive installation by Purin Phanichphant (artist), Benedikt Diemer (astrophysicist), and Natalie Zahr (neuroscientist).

\begin{figure}[H]
    \includegraphics[width=0.307\columnwidth]{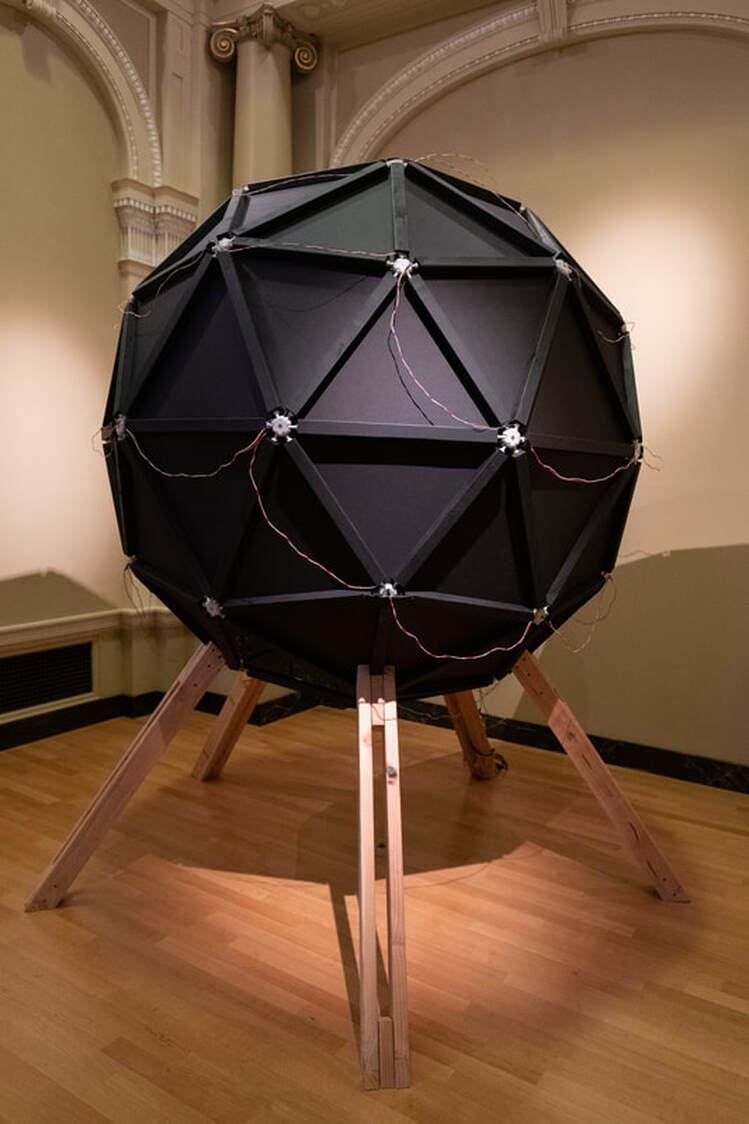}
    \includegraphics[width=0.693\columnwidth]{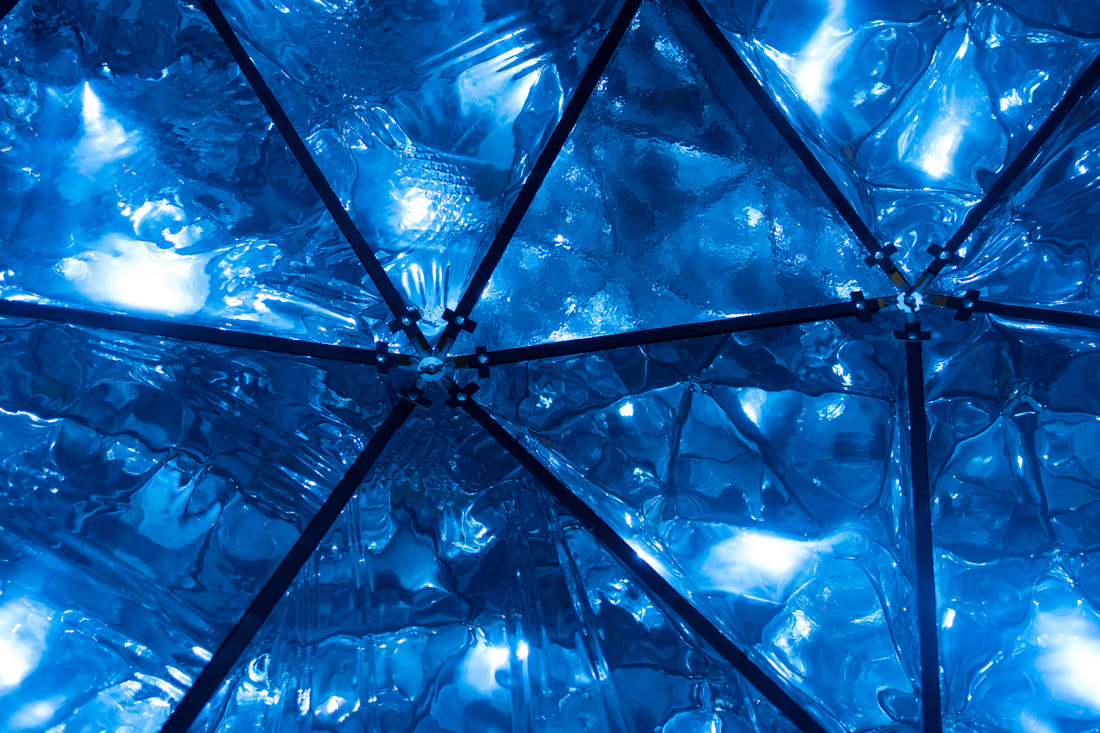}
  \caption{\textit{Chamber of (In)finite Potential}, exterior and interior view. Photo credit Wendy Leyden.}
  \label{chamber}
\end{figure}
\vspace{-1mm}

In this piece, we consider a fundamental similarity between the cosmic web of dark matter in the Universe and the complex connections of the human brain: the Universe has the potential to form a nearly limitless number of structures (such as galaxies); brain cells have the potential to form a nearly limitless number of connections. However, both the Universe and the brain have -- at this point in time -- achieved only a fraction of their potential.
 
Furthermore, both fields of study are in the midst of lively debates as to whether the current understanding fully explains reality. While we tend to think of the Universe as infinite, the growth of structure within it is not: as dark energy accelerates cosmic expansion, gravity will eventually be unable to overcome the stretching, and structures will be frozen. Similarly, while the abundance of neurons in the brain is limited (at birth, the brain has almost all the neurons that it will ever have), the number of connections is potentially infinite; nevertheless, the number of connections reaches a peak and then begins to decay with aging.

These considerations lead to fundamental questions about the nature of the two systems. Will there be a countable number of galaxies in the Universe at the end of its life? Is the capability of our minds fundamentally limited by a finite number of neurons or their connections? These are the kinds of questions we seek to explore in our installation.
 
The seeming contradiction between the infinite and finite nature is expressed through a six-foot wide geodesic dome that is finite and graspable, but with reflective surfaces within that create an infinite visual field. When the viewer enters the dome, his/her perspective is from the center of this perceptual field, metaphorically at the center of the Universe and self-awareness. The artist aims to reveal the infinite potential that exists inside each and every one of us, and that our existence, both in the neural and cosmic realms, are one and the same.

\subsection*{\textit{Natural Science}}

An interactive installation by Thea Boodhoo (scenery design, concept), Cory Bloyd (interactive design, concept), Yagiz Mungan (sound), David Weinberg (astrophysicist), and Dan Feldman (neuroscientist), with contributions by Gary Boodhoo (concept), James Morgan (concept, materials), and Lyn Collier (feltwork).

\begin{figure}[H]
  \begin{center}
      \includegraphics[width=0.7\columnwidth]{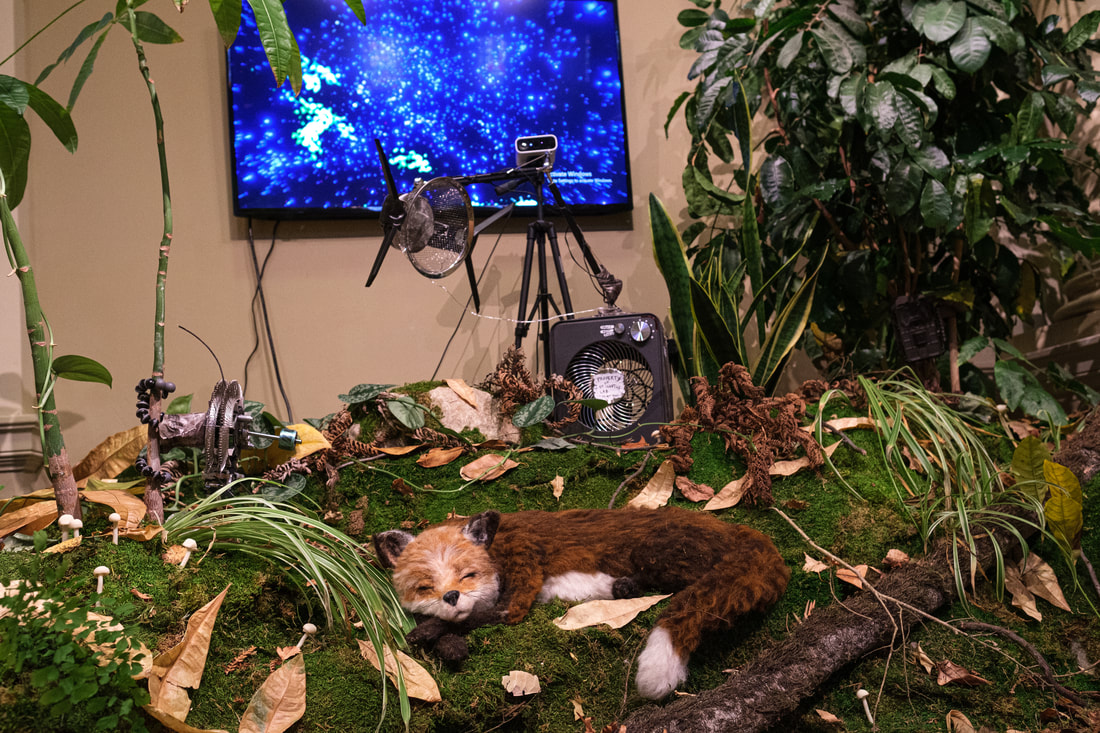}
  \end{center}
  \caption{\textit{Natural Science}, showing fox sleeping with screen above. Photo credit Wendy Leyden.}
  \label{naturalscience}
\end{figure}
\vspace{-1mm}

Natural science, or the scientific study of nature, includes ecology, biology, geology, and physics, as well as cosmology and neuroscience. This surpasses the everyday meaning of nature, going far beyond wildlife and forests to include the farthest reaches of the Universe, as well as the deepest inner workings of the mind.
 
It could be concluded, from this point of view, that all endeavors of natural science are in fact nature studying itself. This piece invites you to broaden your definition of nature, and what you consider natural, dramatically. Don't forget to include yourself.
 
The Science: What do the brain at rest and the cosmos at rest have in common? Science finds that both are surprisingly active. Brain waves cascade from sleeping brains; gravitational waves flood the quietest regions of space. Dan Feldman studies how our neurons process, and often create, what we experience. His ongoing research inspired the setup of our ``experiment,'' as well as a question: could we someday step into another being's dream? What would we encounter? One answer comes from astrophysics. Every brain, and so every dream, is part of the cosmos -- David H. Weinberg's specialty. His research asks: How do galaxies form and cluster? What happens when black holes merge? What was the Universe like before stars, and what will it be like when it finally rests?... Will it dream?
 
The Art: A wild fox has chosen a secluded mossy outcrop as a place to nap before getting back to fox business. Little does he know, researchers studying the inner workings of the mind are running an experiment in this part of the woods. Secret prototype devices tune into the dreams of any animal who sleeps here.
 
The fox's reverie is projected back as sound and vision, and its contents are quite a surprise. Explorers in these woods will hear massive celestial bodies merging and tiny neuron cells firing. If they approach quietly, they may even influence the fox's dream -- for they, like dreams, are part of the cosmos.

\subsection*{\textit{I am a \longlinetwo\ Neuron}}
An interactive installation by Yoon Chung Han (artist), Soo Jung Kwak (artist, sound), Rushi Patel (engineer, software), Jeffrey Kruk (astrophysicist), and Tamira Elul (neuroscientist).

\begin{figure}[H]
  \begin{center}
      \includegraphics[width=0.7\columnwidth]{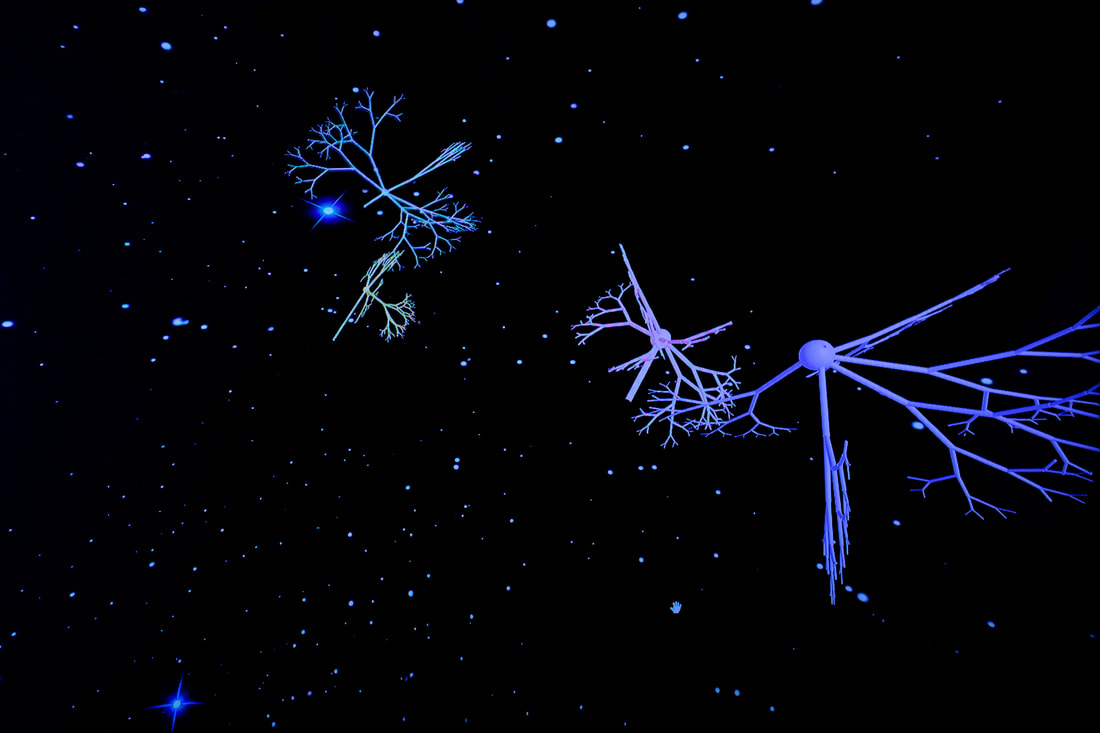}
  \end{center}
  \caption{Screenshot of \textit{I am a \longlineone\ Neuron}.}
  \label{iama1}
\end{figure}

This installation is inspired by the concept of interactions among neighboring galaxies over cosmic time and among neurons during development of the brain. In the forming and expanding Universe, galaxies interact through gravity, winds, and radiation, whereas in the developing brain, neurons are influenced by mechanical, molecular, and electrical cues from other neurons.
 
\textit{I am a \longlinetwo\ Neuron} is an interactive WebVR artwork of creative exploration within a virtual cosmic world. Viewers create their own personalized 3D artificial neurons, interact with them, and observe their interactions. Live feedback from participants controls three factors (gravity, winds, and radiation) to change the environment and neurons. When neurons collide with other neurons, special sounds are emitted, and changes occur in the neurons such as scale, design, opacity, and colors that visually represent interactions of actual neurons in our brains. 

\vspace{-1mm}\begin{figure}[H]
  \begin{center}
      \includegraphics[width=0.7\columnwidth]{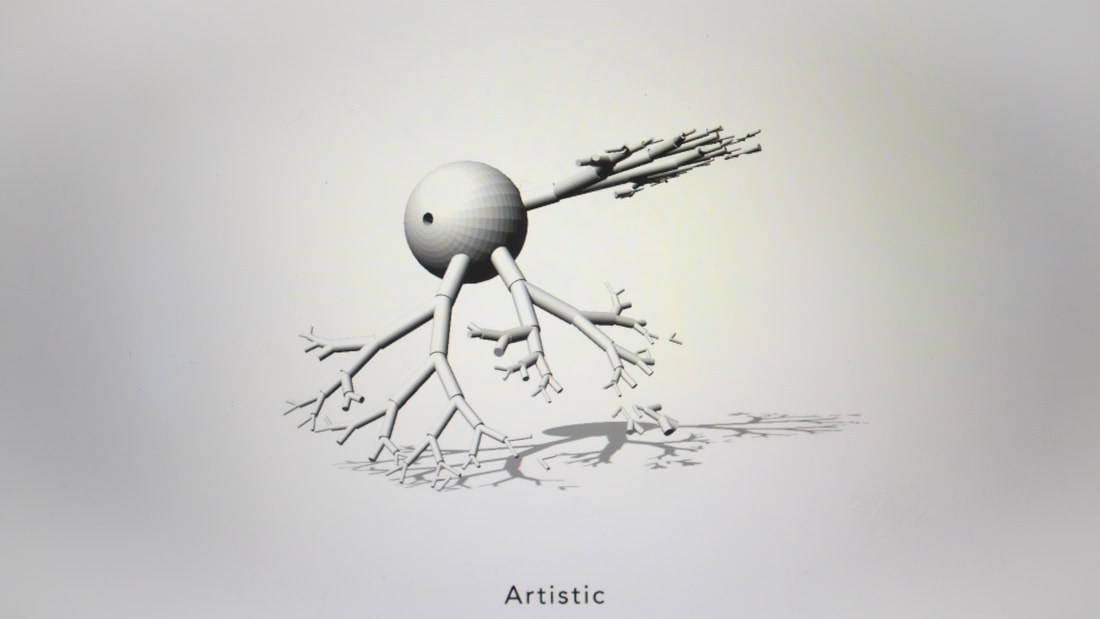}      
  \end{center}
  \caption{\textit{I am a \longlineone\ Neuron} screenshot of an ``artistic'' neutron.}
  \label{iama2}
\end{figure}
\vspace{-1mm}

Through this artwork, participants also explore self-representations that interact with other humans and grow, die, interact, collide, metabolize, reproduce and emit sounds. The participants choose one adjective that represents their personalities for a virtual 3D neuron. Participants relate to these neuron avatars by observing how audiovisual representations change through time. The morphing neurons mimic how we are changed through environmental factors, social groups, and human relationships. Through this artwork participants can find personal meanings into what nature is.
 
One factor contributing to the success of this collaboration was that the scientists trusted the artist to develop an interpretation of this concept independently, without further trying to influence her process beyond the initial concept.

\subsection*{\textit{One in the Universe}}
A virtual reality experience by Anastasia Victor (artist), Mark Neyrinck (astrophysicist), Miguel Angel Arag\'{o}n-Calvo (astrophysicist), and Michael Silver (neuroscientist).

\begin{figure}[H]
  \begin{center}
      \includegraphics[width=0.7\columnwidth]{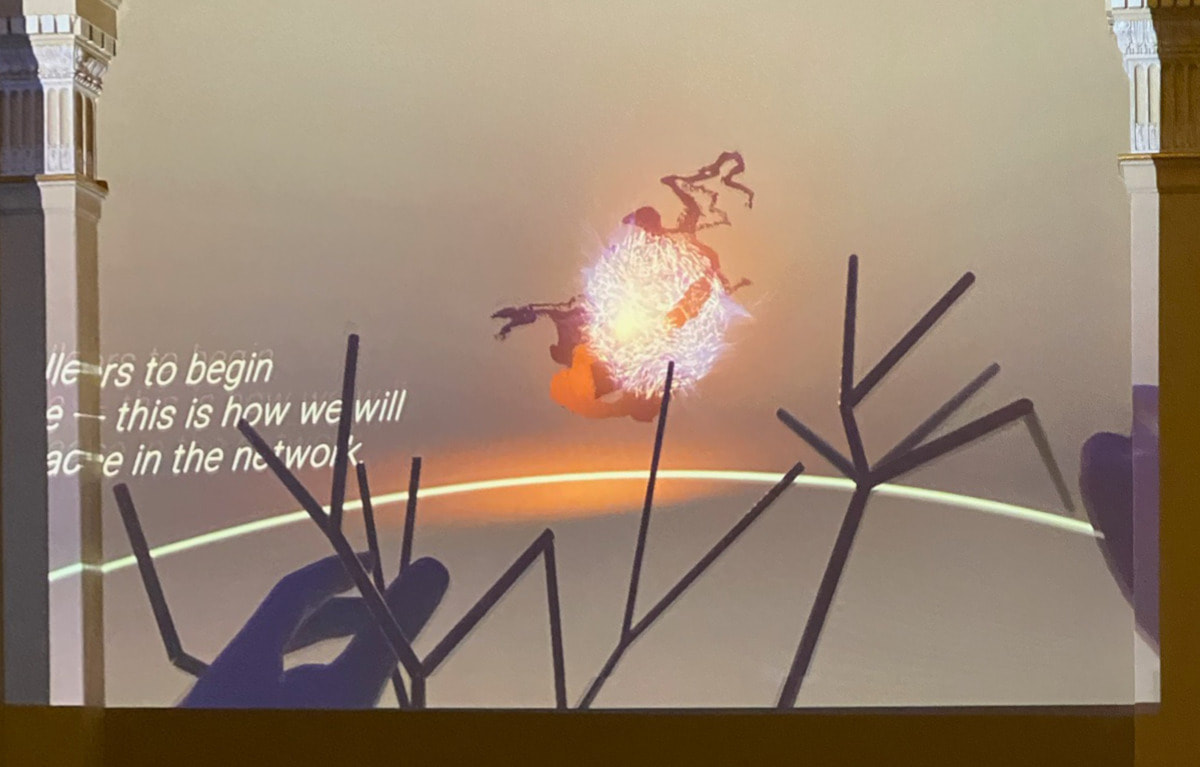}
  \end{center}
  \caption{\textit{One in the Universe} participant constructing their neuron, reaching out to others. Photo credit Wendy Leyden. }
  \label{onein}
\end{figure}
\vspace{-1mm}

Despite vast differences in spatial scale, there is a conceptual link between networks in the brain and those in the cosmic web: they both contain branching structures that connect neurons and nodes of dark matter to one another. Emergent network properties that arise from connectivity of individual elements are also evident for human beings and their relationships. 

\begin{figure}[H]
  \begin{center}
    \includegraphics[width=0.7\columnwidth]{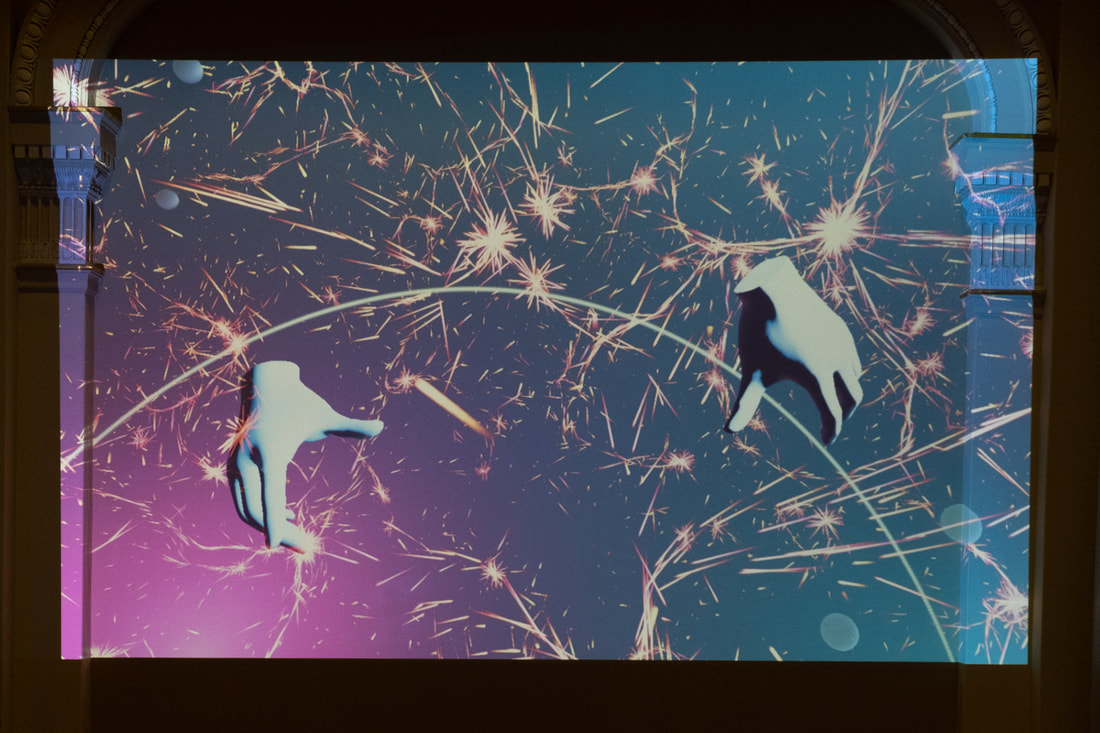}
  \end{center}
  \caption{The full, connected cosmic web from the Bolshoi (\url{http://hipacc.ucsc.edu/Bolshoi.html}) Simulation in \textit{One in the Universe}. Photo credit Wendy Leyden.}
  \label{onein}
\end{figure}
\vspace{-1mm}

\textit{One in the Universe} is an interactive installation that asynchronously connects people across time and space. Within the current climate of filter bubbles and social division, this piece provides moments of introspection and connection to others.

Participants in this installation navigate through a virtual reality composed of structures inspired by dark matter webs and the shapes of neurons, and their movements guide the growth of neural branches within this space. Participants also provide verbal responses to prompts about their own experiences of feeling connected, and these responses are recorded and integrated into the virtual Universe.

The virtual Universe is additive, with each new participant's movements through the space and their verbal responses contributing to the growing patterns of connectivity. Participants directly experience this connectivity by traversing the virtual Universe and hearing verbal responses of previous participants at specific locations in the network. Other visitors to the exhibit observe participants' interactions with the virtual Universe through a projected image that mirrors the visual experience in the virtual reality headset.
 
A scientific project grew out of this piece, as well: we found preliminarily, and cannot yet explain, that the scaling of the total wiring length of the cosmic web with the number of its nodes is similar to the scaling of the wiring length of a neuron with its branch points. They scale roughly with a power law obeyed by a minimal spanning tree, found for neurons by \citet{Cuntz2012}. We came upon this relationship through using the Cuntz et al.\ algorithm to grow branches of the neuron based on input from the participant's hands.

\subsection*{\textit{voi!drum!emory}}
An interactive installation by Miguel Novelo (artist), Bridget Falck (astrophysicist), and Sarah Banducci (neuroscientist).

\begin{figure}[H]
  \begin{center}
      \includegraphics[width=0.7\columnwidth]{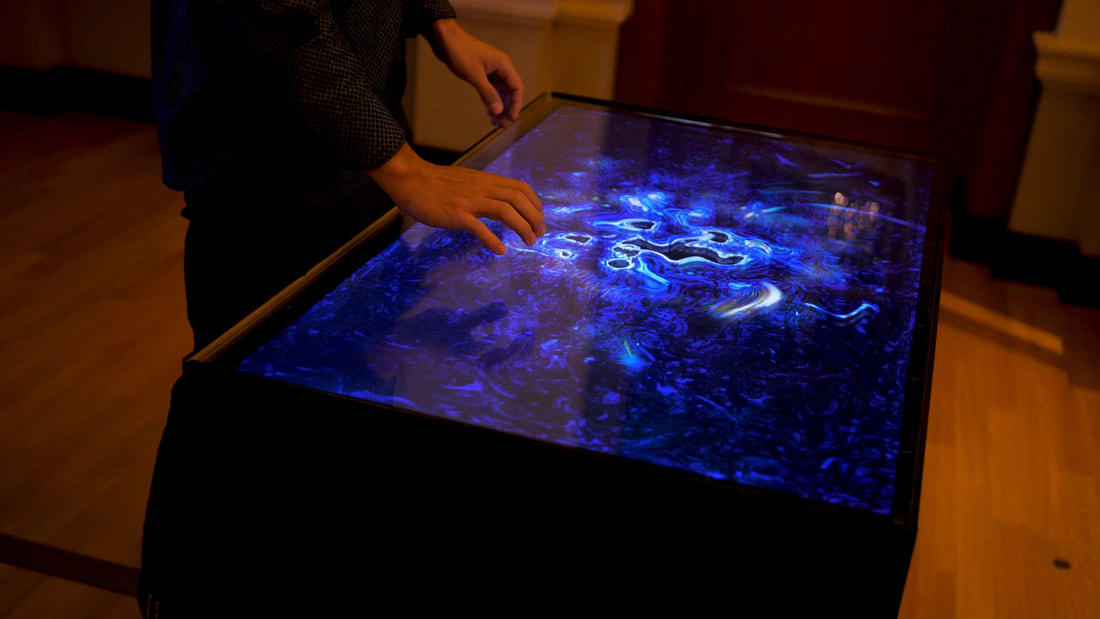}
  \end{center}
  \caption{\textit{voi!drum!emory} close-up, photo credit Wendy Leyden.}
  \label{voidrum2}
\end{figure}
\vspace{-1mm}

A void is not merely an absence, a loss, an emptiness, or a lack, though it can be these things. By calling something a void, we give it a life of its own. It becomes a negative space that defines the positive. In the Universe, voids push matter away, and where they collide with each other, they create the cosmic web, and galaxies. In brains, the web of interconnected neurons deteriorates due to aging and disease; empty space takes over as the brain atrophies. Both types of voids are not static: they reflect the growing, aging, and dying of the Universe and the people that live in it. Our team of a neuroscientist (Sarah Banducci), a cosmologist (Bridget Falck), and an artist (Miguel Novelo) explored this connection between brain and cosmic voids over a year of emails and video calls.

\begin{figure}
  \begin{center}
      \includegraphics[width=0.7\columnwidth]{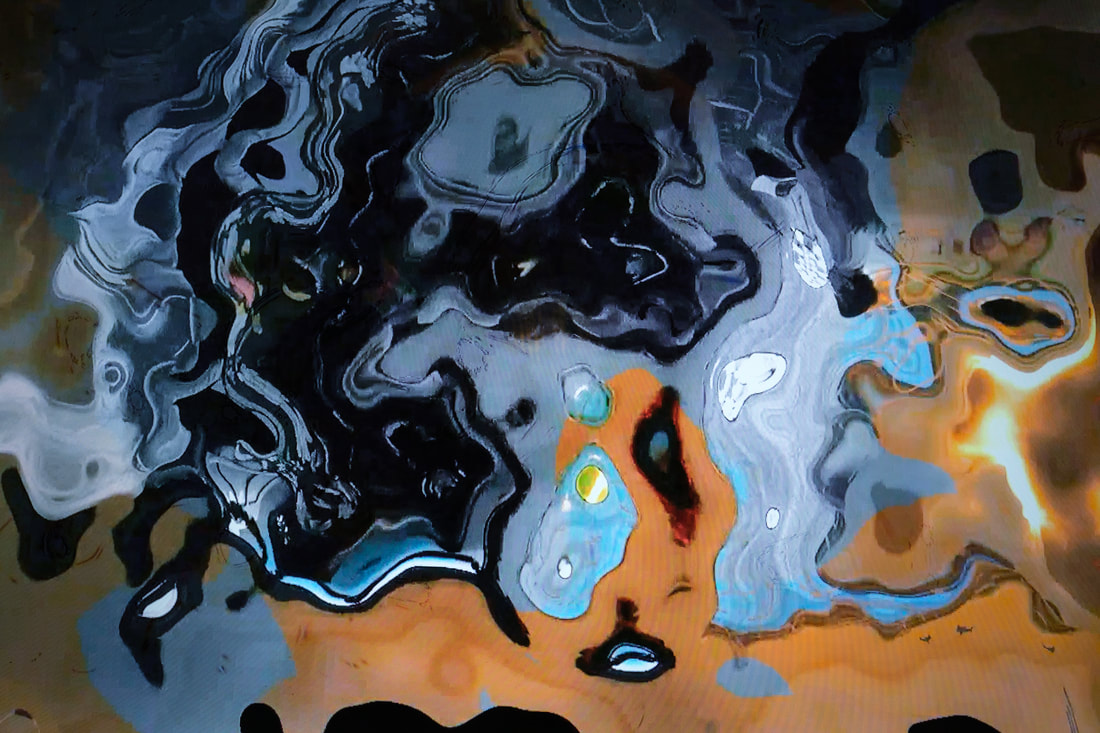}      
  \end{center}
  \caption{\textit{voi!drum!emory} close-up, photo credit Miguel Novelo.}
  \label{voidrum2}
\end{figure}
\vspace{-1mm}

The collaboration kicked off with Falck and Banducci, two strangers on opposite sides of the country, sharing their love of science rooted in two very different fields. Pleasantries aside, both scientists dove past the superficial and into the meat of their respective interests. The first brainstorm ended with a list of four potential ideas for overlap between cosmology and neuroscience: density, connectivity, folds, and voids. By the end of the second call two weeks later, it became clear that voids would provide the greatest opportunity for exploration between the two fields.
 
For Falck, voids are exciting because they are the largest structures in the cosmic web -- except they aren't structures at all. Voids are the emptiest regions of space, and they are growing. As the Universe ages, gravity causes matter to collapse in on itself and pushes matter away from voids, making relatively empty regions of space become even emptier. The aging of the Universe thus grows empty space.

On a much smaller scale, Banducci views voids as a byproduct of human aging. In contrast to the cosmic web, where the force of pulling matter together forms emptiness, in the brain, voids are formed by matter deteriorating. This deterioration of neurons and connections between them gives the appearance of a shrinking brain. Evidence of this atrophy looks different for each individual but translates to characteristic cognitive and motor impairments in older adults.
 
As an artist, Novelo's initial thoughts were to materialize the void with objects that might have a void or vacuum. From an empty cylinder, his sketch led to a percussive instrument. Once a drum is hit on the top membrane, air moves inside and sound waves reverberate in concave space. This sound illustrates a wholeness of the object, the echo, and decay symbolizing the void in a sonic way. Taking the concept a step further, Novelo explored the idea of experiencing a void by sensing the end of reverberation, paying attention to the time between echoes and the sound fading out. This became the physical representation of the void.
 
To add an emotional component to this experience, the team considered, ``What is left after the void?'' Discussing memory in this context, Novelo, Falck, and Banducci came up with poetic voids -- remembrance of what is not there anymore. In the cosmic web, what is left could be the photons finally reaching us from an event that happened millions of years ago; in the brain, the rest of our neurons or gray matter are still intact; and in our memory, previous experiences, nostalgia, and trauma. The art object we conceptualized and created has the characteristic of activating sound and visual cues that symbolize time prior to the void, rapid decay, and lastly a void surrounded by memory.
 
The final expression of this project should leave the viewer with the assurance that even when there is a void, some piece of the experience will continue to linger.

\subsection*{\textit{The Undulating Architecture Illuminating the Individual Sciences}}
A video installation by Forest Stearns (artist), Joel Primack (astrophysicist), and Olaf Sporns (neuroscientist).

\begin{figure}[H]
  \begin{center}
      \includegraphics[width=0.307\columnwidth]{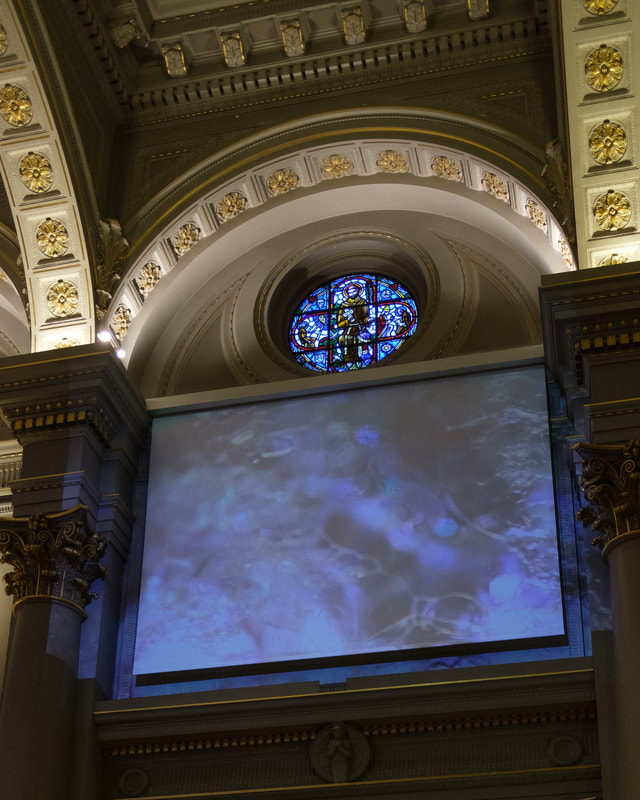}
      \includegraphics[width=0.683\columnwidth]{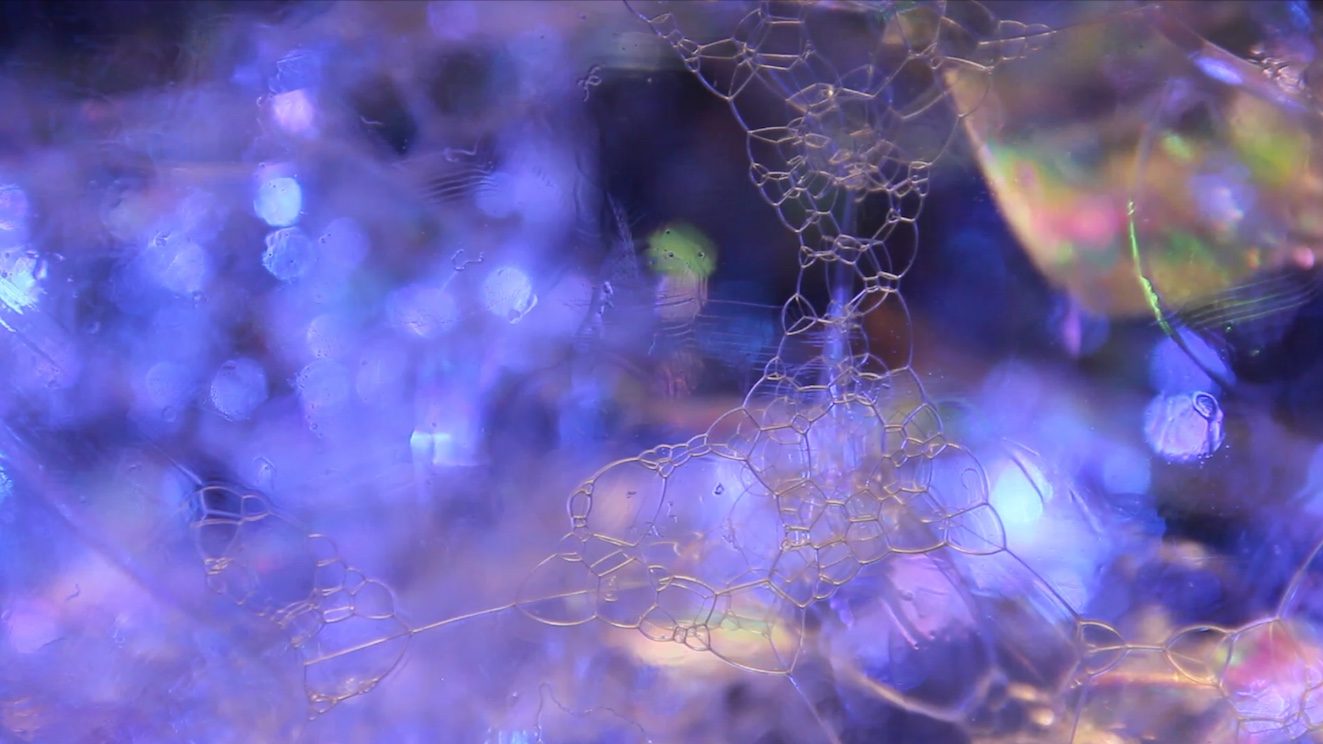}      
  \end{center}
  \caption{\textit{The Undulating Architecture Illuminating the Individual Sciences}. Left: as projected in Manresa Gallery. Right: still image from video; full video at \url{https://vimeo.com/385418604}.}
  \label{undulating}
\end{figure}
\vspace{-1mm}

As a collective, our team of three spent a good amount of time investigating the superficial similarities between dark matter of the universe and the neural networks of our brains. After much discussion, we collectively agreed that the differences greatly superseded the similarities.
 
This being the case, the artist focused on the architectures of both sciences. Using soap bubbles as a medium, the artist found that the undulating bubbles with their walls, filaments, and nodes illuminated the architecture of the universe. The relationships between cosmic webs and soap bubble foams has also been explored by \citet{AragonCalvo2014}. The soap running down filaments towards central nodes and into a bigger pool is similar to how impulses move across the neural net. The physical mass of bubbles were an abstracted screen, projected onto videos that graphically explain the specific sciences. The camera collected the photons rippling across the bubble field.

\section*{Acknowledgments}
MCN is grateful for funding from Basque Government grant IT956-16.

The first three authors were primarily responsible for this article as it appears, but all others contributed ideas and/or text. Curator Esther Mallouh originated the project.

\bibliographystyle{hapj}
\renewcommand{\refname}{\Large References}
\small

\bibliography{refs,refs_res}

\theendnotes

\end{document}